\documentclass[a4paper,10pt, twocolumn, aps, superscriptaddress,notitlepage
 amsmath,amssymb,
 footinbib,
 prb,
floatfix,
longbibliography
]{revtex4-2}
\pdfoutput=1
\usepackage{amsfonts}

\usepackage{bm}
\usepackage{physics}
\usepackage{mathtools}
\usepackage{url}
\usepackage[svgnames,dvipsnames, x11names]{xcolor}
\usepackage{tikz}
\usetikzlibrary{intersections,calc,arrows.meta, decorations.markings}
\usepackage{comment}
\usepackage{lipsum}
\usepackage{adjustbox}

\newtheorem{theorem}{Theorem}
\newcommand{\iu}{\mathrm{i}}

\usepackage[colorlinks,citecolor=DarkGreen,linkcolor=blue,linktocpage]{hyperref}

\begin{document}

\title{Towers of quantum many-body scars from integrable boundary states}
\date{\today}
\author{Kazuyuki Sanada}
\affiliation{Department of Physics, The University of Tokyo, Hongo, Bunkyo-ku, Tokyo 113-0033, Japan}

\author{Yuan Miao}
\affiliation{Kavli Institute for the Physics and Mathematics of the Universe (WPI), The University of Tokyo Institutes for Advanced Study, The University of Tokyo, Kashiwa, Chiba 277-8583, Japan}
\author{Hosho Katsura}
\affiliation{Department of Physics, The University of Tokyo, Hongo, Bunkyo-ku, Tokyo 113-0033, Japan}
\affiliation{Institute for Physics of Intelligence, The University of Tokyo,  
Hongo, Bunkyo-ku, Tokyo 113-0033, Japan}
\affiliation{Trans-scale Quantum Science Institute, The University of Tokyo, Bunkyo-ku, Tokyo 113-0033, Japan}

\begin{abstract}
We construct several models with multiple quantum many-body scars~(QMBS) using integrable boundary states~(IBS). Specifically, we focus on the tilted Néel states, which are parametrized IBS for the spin-1/2 Heisenberg chain, and show that these states can be used to construct a tower of scar states. Our models exhibit periodic revival dynamics, showcasing a characteristic behavior of superpositions of QMBS. Furthermore, the tower of QMBS found in this study possesses a restricted spectrum generating algebra~(RSGA) structure, indicating that QMBS are equally spaced in energy. This approach can be extended to two-dimensional models, which can be decomposed into an array of one-dimensional models. In this case, the tilted Néel states again serve as parent states for multiple scar states. These states demonstrate low entanglement entropy, marking them as exact scar states. Notably, their entanglement entropy adheres to the sub-volume law, further solidifying the nonthermal properties of QMBS. Our results provide novel insights into constructing QMBS using IBS, thereby illuminating the connection between QMBS and integrable models.
\end{abstract}
\maketitle

\section{Introduction}
Thermalization in isolated quantum many-body systems is a long-standing puzzle in the study of quantum physics. Initiated by von Neumann's seminal work on the quantum ergodic theorem~\cite{neumann1929beweis}, there have been numerous theoretical studies on quantum thermalization~\cite{ekstein1957ergodic, berry1977regular}. Recently, the eigenstate thermalization hypothesis (ETH)~\cite{deutsch1991quantum, srednicki1994chaos, Deutsch_2018} has been recognized as a central principle that governs quantum thermalization. ETH provides a framework for explaining the relaxation of quantum many-body systems towards equilibrium. Roughly speaking, the strong version of ETH states that all energy eigenstates are thermal. It has been demonstrated that the strong ETH holds for many types of quantum systems~\cite{rigol2008thermalization, polkovnikov2011colloquium, nandkishore2015many, D_Alessio_2016}. However, it does not apply to all types of quantum many-body systems; 
known exceptions include integrable systems~\cite{Ikeda_PRB_2013, Alba_PRB2015, brenes2020eigenstate, brenes2020low}, many-body localized systems~\cite{nandkishore2015many, abanin2017recent, abanin2019colloquium, fabien2018manybody}, and systems with Hilbert space fragmentation~\cite{moudgalya2022quantum, moudgalya2022hilbert, brighi2023hilbert,Buca_PRX2023}.

Recent advances in experimental techniques, such as quantum simulators with superconducting circuits~\cite{neill2016ergodic, morvan2022formation}, trapped ions~\cite{smith2016many}, ultracold atoms~\cite{trotzky2012probing, langen2015ultracold, ueda2020quantum}, and Rydberg atoms~\cite{bernien2017probing, adams2019rydberg}, have made it possible to directly observe thermalization in quantum systems. This development revealed an exotic class of quantum many-body states that evade thermalization~\cite{bernien2017probing}. These nonthermal states are called quantum many-body scars (QMBS), in analogy to quantum scars in single-particle systems~\cite{heller1984bound, Kaplan_1998, turner2018weak}. 
Today, it is widely accepted that QMBS refer to energy eigenstates of non-integrable models that violate ETH~\cite{serbyn2021quantum}. To date, a number of systems exhibiting QMBS have been reported in both experimental~\cite{su2023observation, zhang2023many, austin2024observation} and theoretical studies~\cite{Turner2, schecter2019weak, choi2019emergent, lin2019exact, Iadecola2020, langlett2021hilbert, serbyn2021quantum, moudgalya2022quantum, papic2022weak, chandran2023quantum, Klebanov_2020, Klebanov_2021, Klebanov_2023, kunimi2024proposal}. 
Some of these systems possess exact QMBS, which can be written in closed form. A systematic construction of such models has become a key issue in theoretical studies of QMBS~\cite{shiraishi2017systematic, o2020tunnels, shibata2020onsager, moudgalya2020large, mark2020eta, Klebanov_2020, mcclarty2020disorder, ren2021quasisymmetry, moudgalya2022quantum, ren2022deformed, chandran2023quantum, omiya2023fractionalization}.  

In our previous work~\cite{sanada2023quantum}, we proposed a method for constructing models with QMBS using integrable boundary states (IBS)~\cite{piroli2017integrable}. These states are associated with a set of conserved charges $\{Q_n\}_{n = 2, 3, \ldots}$~\footnote{The subscript $n$ of the charge $Q_n$ denotes its \textit{order}, which indicates how many consecutive sites its local operator spans.} in quantum integrable models. An IBS is defined as a state that is annihilated by all parity-odd charges of an integrable Hamiltonian~\cite{piroli2017integrable, de2018scalar, Pozsgay2018overlaps, piroli2019integrable, pozsgay2019integrable}, i.e., $\{Q_{2k+1}\}_{k = 1, 2, \ldots}$. In other words, an IBS $\ket{\Psi_0}$ is a state for which
\begin{equation}
    Q_{2k+1}\ket{\Psi_0} = 0 
\end{equation}
holds for all $k = 1, 2, \ldots$.

The IBS play an important role in the study of out-of-equilibrium physics of quantum integrable models \cite{caux2014quench, takacz2014quench, caux2016quench, piroli2017integrable}, as well as in obtaining correlation functions in the context of AdS/CFT integrability \cite{deLeeuw2015onepoint, zarembo2016MPS, foda2016overlaps, de2018scalar, Linardopoulos2020}. The IBS under consideration are typically quantum states with area-law entanglement entropy scaling, such as product states or matrix product states~\footnote{There is one notable exception with volume-law entanglement scaling, the crosscap state \cite{komatsu2022crosscap, gombor2022crosscap, ekman2022crosscap, jiang2023crosscap, chiba2024exact}, which is out of this article's scope.}. One of the remarkable examples is the tilted Néel states for the isotropic and anisotropic Heisenberg spin chains. We will use the tilted Néel states in our paper to construct QMBS.

Our procedure starts with a given IBS $\ket{\Psi_0}$. 
We identify a nonintegrable operator $H_\mathrm{NI}$ for which $\ket{\Psi_0}$ is an eigenstate~\footnote{The method in Ref. \cite{Qi2019determininglocal} is particularly useful for finding such an operator.}. If such a Hamiltonian $H_\mathrm{NI}$ is found, $\ket{\Psi_0}$ becomes an energy eigenstate of 
\begin{equation}
    H(\{t_k\}) = H_\mathrm{NI} + \sum_{k=1}^\infty t_k Q_{2k+1},
\end{equation}
where $\{t_k\}^\infty_{k=1}$ are real numbers. The state $\ket{\Psi_0}$ can be regarded as an exact QMBS if the Hamiltonian $H(\{t_k\})$ is nonintegrable and the energy of this state lies in the middle of the spectrum. 
In our previous work \cite{sanada2023quantum}, we demonstrated the effectiveness of this approach by constructing several families of models with QMBS, e.g., the Majumdar-Ghosh model perturbed by parity-odd charges of the Heisenberg chain. 
However, the models constructed by this method exhibit only one or two isolated QMBS that do not have any nontrivial dynamics. 
In this paper, we construct models with multiple QMBS using IBS and show that their dynamics exhibit periodic revivals, which is a typical behavior for a superposition of QMBS~\cite{alhambra2020revival}.

This paper is organized as follows. In Sec. \ref{sec:XXX_based}, we introduce a model with QMBS that are annihilated by the scalar spin chirality, the third conserved charge of the spin-$1/2$ XXX model. In Sec. \ref{sec:XYZ}, we demonstrate that the model introduced in Sec. \ref{sec:XXX_based} can be generalized to the completely anisotropic case, where the scalar spin chirality is replaced with the third conserved charge of the spin-$1/2$ XYZ model. 
In this generalized model, the IBS discussed in Sec. \ref{sec:XXX_based} remains a parent state of QMBS in the new model. In Sec. \ref{sec:dynamics}, we examine the dynamics of a superposition of scar states and compare it with that of a thermal state. In Sec. \ref{sec:higher-dimensional}, we discuss a higher-dimensional extension of our models introduced in Sec. \ref{sec:XYZ}.
Finally, we conclude and discuss our results in Sec. \ref{sec:discussions}.

\section{Model with U(\texorpdfstring{$1$}{1}) symmetry and scars}
\label{sec:XXX_based}
We consider a one-dimensional (1D) spin-$1/2$ model with two- and three-body interactions. The Hamiltonian depends on two parameters $g$ and $h_y \in \mathbb{R}$ and is given by
\begin{equation}
    H_1(g, h_y) = C_\mathrm{SC} + g H_\mathrm{pert} + h_y Y, \label{eq:ham}
\end{equation}
where
\begin{align}
    C_\mathrm{SC} &= \sum_{j=1}^L \bm{\sigma}_j\cdot(\bm{\sigma}_{j+1}\times \bm{\sigma}_{j+2}), \label{eq:CSC} \\
    H_\mathrm{pert} &= \sum_{j=1}^Lc_j(\sigma_j^x\sigma_{j+1}^x-\sigma_j^y\sigma_{j+1}^y+\sigma_j^z\sigma_{j+1}^z+1), \label{eq:Hpert} \\
    Y &= \sum_{j=1}^L \sigma_j^y. \label{eq:Yop}
\end{align}
Here $\sigma^\mu_j$ ($\mu=x,y,z$) are the Pauli matrices acting on site $j$ and each of $\{c_j\}^L_{j=1}$ can be any real number. We impose periodic boundary conditions, i.e., $\sigma^\mu_{L+1}=\sigma^\mu_1$, and assume the number of sites $L$ to be even.

The only symmetry of the Hamiltonian $H_1$ is ${\rm U}(1)$ symmetry generated by $Y$, which is the $y$-component of the total spin. Because of this symmetry, the Hilbert space splits into different sectors labeled by the eigenvalues of $Y$. In the following, with a slight abuse of notation, we will denote the eigenvalue of $Y$ by the same symbol. 

\begin{figure}
    \centering
    \includegraphics[width = \linewidth]{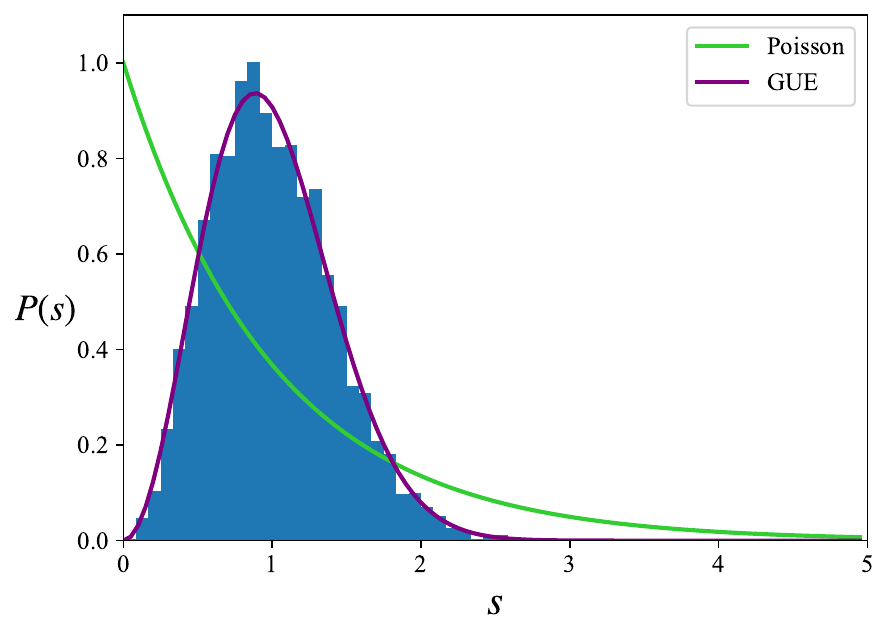}
    \caption{Level-spacing statistics in the middle half of the spectrum of $H_1(g, h_y)$ [Eq. \eqref{eq:ham}] with $g = 1$, $h_y = 1$, and $L=16$. The coefficients $c_j$ are randomly chosen from $[-1, 1]$. The data are taken in the symmetry sector with $Y = 0$. The Wigner-Dyson (GUE) and Poisson distributions are shown for comparison.}
    \label{fig:xxz_leveldist}
\end{figure}

The Hamiltonian $H_1$ is unlikely to be integrable since the perturbation term $H_\mathrm{pert}$ is nonintegrable for generic coefficients $\{c_j\}^L_{j=1}$. 
To confirm the lack of integrability of the Hamiltonian $H_1$, we calculate the distribution $P(s)$ of the normalized energy level spacings $s_i=(E_i-E_{i-1})/\Delta$, where $\{E_i\}$ is the set of energy eigenvalues sorted in ascending order and $\Delta$ denotes the mean level spacing. 
According to random matrix theory, the level-spacing distribution obeys the Wigner-Dyson distribution 
\begin{equation}
    P_\mathrm{WD}(s) = \frac{32}{\pi^2}s^2e^{-\frac{4}{\pi}s^2}
\end{equation}
if the system is nonintegrable~\footnote{More precisely, the level-spacing distribution obeys $P_\mathrm{WD}(s)$ when the given system does not have time-reversal symmetry. When the system is nonintegrable and time-reversal invariant, its level spacing distribution obeys the Gaussian $\beta$ ensemble with $\beta=1$, i.e., $P(s) = \frac{\pi}{2}se^{-\frac{\pi}{4}s^2}$.}~\cite{bohigas1984characterization,kundu2023signatures}, and it obeys the Poisson distribution
\begin{equation}
    P_\mathrm{Poisson}(s) = e^{-s}
\end{equation}
if the system is integrable or many-body localized~\cite{buijsman2019random}. 
Figure \ref{fig:xxz_leveldist} shows that $P(s)$ of the energy spectrum of $H_1$ in the sector with $Y=0$ obeys the GUE Wigner-Dyson distribution rather than the Poisson distribution.
This suggests that the Hamiltonian is neither integrable nor in a many-body localized phase.

Having established the non-integrability of the model, we now turn to the construction of QMBS. Our target states are the tilted N\'{e}el states parametrized by $\alpha \in \mathbb{C}$:
\begin{align}
    \ket{\psi_1(\alpha)} &= \frac{1}{\sqrt{\mathcal{N}}}\left[(-\alpha\ket{\uparrow}+\ket{\downarrow})\otimes(\ket{\uparrow}+\alpha\ket{\downarrow})\right]^{\otimes \frac{L}{2}}, \label{eq:psi1}\\
    \ket{\psi_2(\alpha)} &= \frac{1}{\sqrt{\mathcal{N}}}\left[(\ket{\uparrow}+\alpha\ket{\downarrow})\otimes(-\alpha\ket{\uparrow}+\ket{\downarrow})\right]^{\otimes\frac{L}{2}},
    \label{eq:psi2}
\end{align}
where $\ket{\uparrow}$ and $\ket{\downarrow}$ are normalized eigenstates of $\sigma^z$ with eigenvalues $+1$ and $-1$, respectively. Here, the parameter $\alpha$ determines the tilt angle and the relative phase of the local spins. The normalization constant is given by $\mathcal{N} = (\abs{\alpha}^2+1)^{L}$. These states are related by one-site translation $\mathcal{T}$.

Since the tilted Néel states $\ket{\psi_{1, 2}(\alpha)}$ are IBS of the 1D Heisenberg model~\cite{piroli2017quantum, piroli2017integrable}, they are annihilated by $C_\mathrm{SC}$, known as the third conserved charge of the Heisenberg Hamiltonian~\cite{Grabowski_1994, Frahm1997properties}. Meanwhile, $H_\mathrm{pert}$ is a weighted sum of projection operators onto $\frac{1}{\sqrt{2}}(\ket{\uparrow\uparrow}_{j, j+1}+\ket{\downarrow\downarrow}_{j, j+1})$, each of which also annihilates the states $\ket{\psi_{1, 2}(\alpha)}$. 
Since $\ket{\psi_{1, 2}(\alpha)}$ are annihilated by $C_{\rm SC}$ and $H_{\rm pert}$ simultaneously, they are zero-energy states of $C_{\rm SC}+g H_{\rm pert}$ for all $g$.
We note in passing that the local terms of $H_\mathrm{pert}$ form a representation of the periodic Temperley--Lieb algebra \cite{temperley1971relations, pasquier1990qg}. 
This is not surprising, since each local term of $H_\mathrm{pert}$ is related to the usual isotropic Heisenberg Hamiltonian on two sites by a unitary transformation.

The key to constructing multiple scar states is to note that the parameter $\alpha$ in the tilted Néel states is arbitrary. This allows us to project the tilted Néel states onto states with definite magnetization in the $y$ direction, which remain eigenstates of $H_1(g,h_y)$ with nonzero $h_y$. To see this, we change the basis from $\{\ket{\uparrow}, \ket{\downarrow}\}$ to the basis in which $\sigma^y$ is diagonal: 
\begin{equation}
    \ket{+} = \frac{1}{\sqrt{2}}(\ket{\uparrow} + \mathrm{i}\ket{\downarrow}), \quad \ket{-} = \frac{1}{\sqrt{2}}(\ket{\uparrow} - \mathrm{i}\ket{\downarrow}).
\end{equation}
In the new basis, $\ket{\psi_1(\alpha)}$ can be expressed as
\begin{equation}
\begin{split}
    \ket{\psi_1(\alpha)}
    = \left(\frac{\alpha-\iu}{\sqrt{2}}\right)^L \cdot \frac{\iu^{L/2}}{\sqrt{\mathcal{N}}} & [(z\ket{+}+\ket{-})\\
    & \otimes (z\ket{+}-\ket{-})]^{\otimes \frac{L}{2}}  \\
    = \left(\frac{\alpha-\iu}{\sqrt{2}}\right)^L\cdot\frac{\iu^{L/2}}{\sqrt{\mathcal{N}}}& \sum_{n=0}^Lz^{L-n} \sqrt{\begin{pmatrix} L \\ n \end{pmatrix}} \ket{{\Psi}_n},
\end{split}
\end{equation}
where $z=\frac{\alpha+\mathrm{i}}{\alpha-\mathrm{i}}$ and
\begin{equation}
\ket{{\Psi}_n} = \frac{({\mathcal{O}}_\pi^-)^n}{\sqrt{n !L!/ (L-n)!}} \ket{++\cdots +}
\label{eq:psin}
\end{equation}
with
\begin{equation}
    {\mathcal{O}}_\pi^- = \sum_{j=1}^L (-1)^{j-1} \Tilde{\sigma}_j^- = \sum_{j=1}^L \frac{(-1)^{j-1}}{2} (\sigma_j^z-\mathrm{i}\sigma_j^x).
\end{equation}
The states $|\Psi_n \rangle$ are normalized to unity.

\begin{figure}[bpt]
    \centering
    \includegraphics[width=\linewidth]{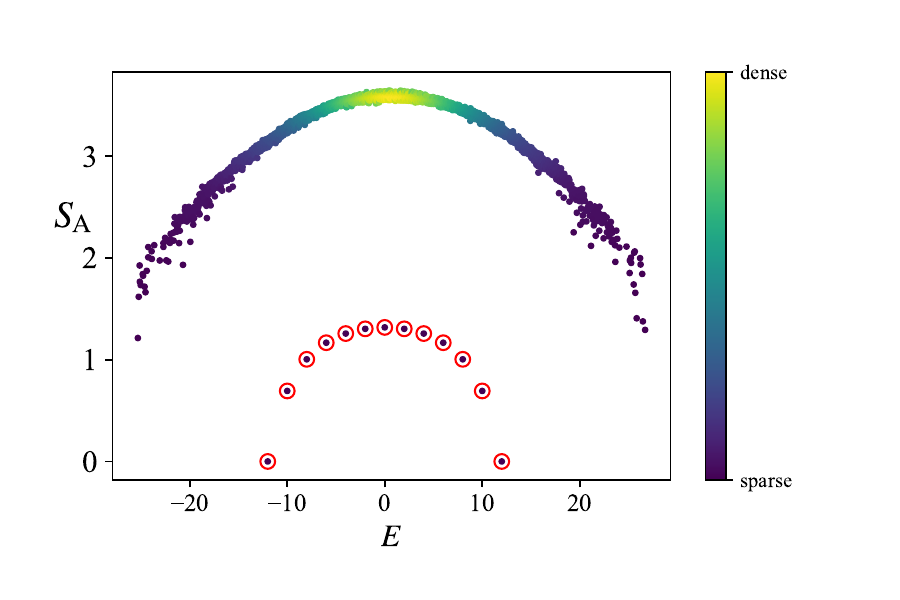}
    \caption{Half-chain EE $S_A$  
    as a function of energy $E$ for all eigenstates of $H_1(g=1, h_y=1)$ with $L=12$. The random coefficients $c_j$ are chosen from the interval $[-1, 1]$. 
    The density of data points is color coded. 
    The red solid circles indicate the scar states $\ket{{\Psi}_n}$ ($n=0,1,...,12$).}
    \label{fig:red}
\end{figure}

Since we can take $z\in \mathbb{C}\setminus \{1\}$ arbitrarily, each $\ket{{\Psi}_n}$ is an eigenstate of the Hamiltonian (\ref{eq:ham}) with eigenenergy $(L-2n)h_y$. To confirm that these states are low-entanglement states in the middle of the spectrum, we compute the half-chain entanglement entropies (EE), $S_A$, of all energy eigenstates. The results are shown in Fig. \ref{fig:red}. Clearly, the half-chain EE of $\ket{{\Psi}_n}$ are significantly lower than those of the other eigenstates, suggesting that these states are exact QMBS. 
We can analytically compute the half-chain EE of $\ket{\Psi_n}$ following the procedure in our previous work~\cite{sanada2023quantum} (see also~\cite{popkov2005logarithmic}). The result reads
\begin{equation}
    S_A(\ket{\Psi_n}) = -\sum_{k=0}^{\min({n, L-n})}\frac{\binom{L/2}{k}\binom{L/2}{n-k}}{ \binom{L}{n}}\ln \frac{\binom{L/2}{k}\binom{L/2}{n-k}}{\binom{L}{n}}.
\end{equation}
For large $L$ at fixed $\frac{n}{L}$, the leading term in $S_A(\ket{\Psi_n})$ is $\frac{1}{2}\ln L$, which proves that their entanglement entropy obeys a sub-volume law.

We note that the scar states $\ket{{\Psi}_n}$ cannot be distinguished from the rest of the states by symmetry, as $[C_\mathrm{SC}, {\mathcal{O}}_\pi^-]\neq 0$. 
However, the presence of a tower of scars can be understood through the restricted spectrum generating algebra (RSGA)~\cite{mark2020unified, moudgalya2020eta}. 
While we focus on isolated quantum systems, we mention that the RSGA framework has also been generalized to open quantum systems in Ref. \cite{buvca2019non}. 
In our system, the Hamiltonian $H_1 (g, h_y)$, the reference state $\ket{\Psi_0}$, and the lowering operator $\mathcal{O}_\pi^-$ satisfy an RSGA of order $2$: 
\begin{align}
    &{\rm (i)}~ H_1 \ket{\Psi_0} = L h_y \ket{\Psi_0}, \\
    &{\rm (ii)}~ [\, H_1, \mathcal{O}_\pi^- \,] \ket{\Psi_0} = \mathcal{E} \mathcal{O}_\pi^- \ket{\Psi_0}, \\
    &{\rm (iii)}~ [\, [\, H_1, \mathcal{O}_\pi^- \,], \mathcal{O}_\pi^- \,] \ket{\Psi_0} =0, \\
    &{\rm (iv)}~ [\, [\, H_1, \mathcal{O}_\pi^- \,], \mathcal{O}_\pi^- \,], \mathcal{O}_\pi^- \,] =0,
\end{align}
where $\mathcal{E}=-2 h_y$. The above relations readily imply that the states $\ket{{\Psi}_n}$ ($n=0,1,2,...,L$) are exact eigenstates of $H_1(g, h_y)$ with eigenvalues $(L-2n)h_y$.

We remark that, beyond the specific Hamiltonian we have discussed so far, one can construct a broader class of models hosting a tower of QMBS by incorporating higher-order conserved charges $Q_{2k+1}$ ($k \ge 2$) of the Heisenberg chain. The key to this construction is that these charges also annihilate the tilted Néel states, as they are IBS; consequently, these charges can be added to the Hamiltonian $H_1(g, h_y)$ with arbitrary real coefficients, while leaving the tower states in Eq. \eqref{eq:psin} as exact eigenstates. See Refs. \cite{Grabowski_1994,grabowski1995structure} for the explicit expressions of $Q_{2k+1}$ ($k \ge 2$). 
Furthermore, since the local terms comprising $H_{\rm pert}$ also annihilate the tilted Néel states, one may include any polynomial in these terms with real coefficients.

\section{Model without U(\texorpdfstring{$1$}{1}) symmetry and scars}
\label{sec:XYZ}
The model constructed in the previous section has a $U(1)$ symmetry corresponding to the spin rotation symmetry about the $y$ axis. In this section, we consider an extension of the model in which this $U(1)$ symmetry is broken explicitly.

Again, our starting point is the tilted N\'{e}el states. One can prove that they are IBS of the XYZ model:
\begin{equation}
\begin{split}
    H_\mathrm{XYZ} &= \sum_{j=1}^Lh_{j, j+1}^\mathrm{XYZ}  \\
    &= \sum_{j=1}^L (J_x\sigma_j^x\sigma_{j+1}^x+J_y\sigma_j^y\sigma_{j+1}^y+J_z\sigma_j^z\sigma_{j+1}^z).
\end{split}
    \label{eq:8vHam}
\end{equation}
(See Appendix~\ref{appendix:NeelIBS} for a proof.) 
Thus, it follows that the third conserved charge $Q_3$ of the XYZ model annihilates $\ket{\psi_{1,2}(\alpha)}$ in Eqs. (\ref{eq:psi1}) and (\ref{eq:psi2}). 
Therefore, $\ket{{\Psi}_n}$ in Eq. (\ref{eq:psin}) are eigenstates of the following Hamiltonian:
\begin{equation}\label{eq:XYZ_H}
    H_2(g , h_y) = Q_3 + g H_\mathrm{pert} + h_y Y,
\end{equation}
with eigenvalues $(L-2n) h_y$, where $H_{\rm pert}$ and $Y$ are defined in Eqs. \eqref{eq:Hpert} and \eqref{eq:Yop}, respectively.

Using the boost operator $B$~\cite{sklyanin1992quantum, Grabowski_1994, grabowski1995structure, deLeeuw2019classifying}, one can explicitly construct the third conserved charge of the XYZ model via $Q_3 = \mathrm{i} [B, H_{\rm XYZ}]$~\footnote{Alternatively, higher order local conserved quantities can be obtained using the method in \cite{Fukai_2020}}. As a result, we obtain \cite{grabowski1995structure}
\begin{equation}
\begin{split}
    Q_3 &= \sum_{j=1}^L \Hat{\bm{\sigma}}_j\cdot(\Bar{\bm{\sigma}}_{j+1}\times\Hat{\bm{\sigma}}_{j+2}) \\
    &= \sum_{j=1}^L \sum_{\lambda, \mu, \nu}\epsilon_{\lambda\mu\nu} \frac{J_xJ_yJ_z}{J_\mu}\sigma_j^\lambda\sigma_{j+1}^\mu\sigma_{j+2}^\nu,
    \label{eq:XYZQ3}
\end{split}
\end{equation}
where
\begin{align}
    \Hat{\sigma}^\mu_j &= \sqrt{J_\mu}\sigma_j^\mu, \label{eq:hatspin} \\
    \Bar{\sigma}^\mu_j &= \sqrt{\frac{J_xJ_yJ_z}{J_\mu}}\sigma_j^\mu, \label{eq:barspin}
\end{align}
and $\epsilon_{\lambda\mu\nu}$ is the Levi-Civita symbol with $\epsilon_{xyz}=1$. Clearly, $Q_3$ reduces to $C_{\rm SC}$ in Eq. (\ref{eq:CSC}) when $J_x = J_y = J_z = 1$. It should be noted that the Hamiltonian $H_2(g, h_y)$ breaks $U(1)$ spin rotation symmetry about the $y$-axis when $J_x \ne J_z$.

\begin{figure}
    \centering
    \includegraphics[width=0.91\linewidth]{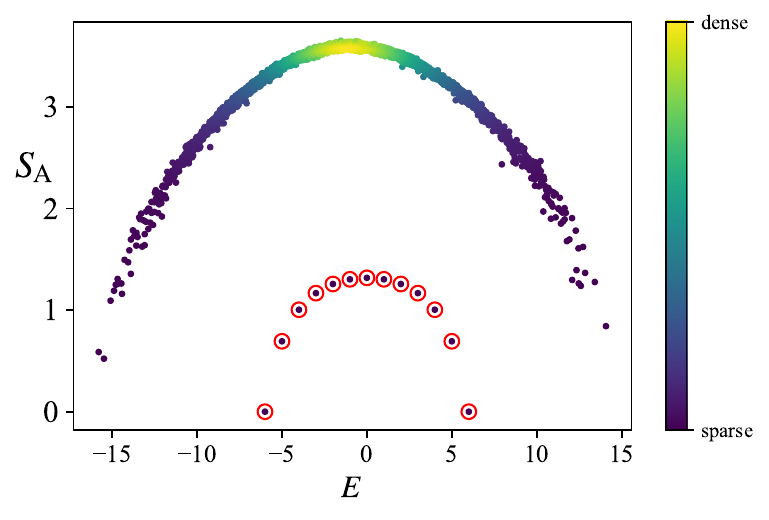}
    \caption{Half-chain EE $S_A$ as a function of energy $E$ 
    for all eigenstates of $H_2(g =1.0, h_y = 0.5)$ with $L=12$ and $(J_x, J_y, J_z) = (0.1, 0.5, 1.0)$. The random coefficients $c_j$ are drawn from the interval $[-1, 1]$. The density of data points is color coded.
    The red solid circles indicate the half-chain EE and eigenenergy of $\ket{{\Psi}_n}$ ($n=0,1,...,12$).}
    \label{fig:XYZpertL12t1y0.5}
\end{figure}

To confirm that $\ket{{\Psi}_n}$ are low-entanglement states in the middle of the spectrum, we compute the half-chain EE for all energy eigenstates of the Hamiltonian \eqref{eq:XYZ_H}. Figure \ref{fig:XYZpertL12t1y0.5} shows the numerical results, where the states $\ket{{\Psi}_n}$ are in the middle of the spectrum and stand out as entanglement outliers with significantly lower entanglement than the other states. This observation leads us to conclude that they are QMBS of the Hamiltonian (\ref{eq:XYZ_H}).

We remark that the tower of QMBS in $H_2(g,h_y)$ can also be explained by an RSGA of order $2$ when $J_x = J_z$, where $\mathrm{U}(1)$ symmetry is manifest. However, when $J_x \ne J_z$, $H_2(g,h_y)$ does not admit this RSGA. Instead, the Hamiltonian exhibits an RSGA of order $4$. We have checked this numerically for $L\le 12$ for various values of the parameters. 
We also remark that higher-order conserved charges $Q_{2k+1}$ ($k \ge 2$) can be added to the Hamiltonian as well, while preserving the tower of QMBS as exact eigenstates.

\section{Dynamics} \label{sec:dynamics}
To reveal the nonthermal features of the scar states, we study their dynamical behavior. Here, we consider the quench dynamics of $H_2$ in Eq. \eqref{eq:XYZ_H}. 
The initial states we consider are the N\'{e}el state $\ket{\mathbb{Z}_2}  = \ket{\uparrow\downarrow\uparrow\downarrow\cdots}$ and a period-3 state $\ket{\mathbb{Z}_3} = \ket{\uparrow\uparrow\downarrow\uparrow\uparrow\downarrow\cdots}$. 
Remarkably, the N\'{e}el state can be written as a weighted superposition of QMBS exactly:
\begin{equation}
\begin{split}
\ket{\mathbb{Z}_2} & = \ket{\uparrow\downarrow\uparrow\downarrow\cdots}=\ket{\psi_1(\alpha \to \infty)} \\
& = \left( -\frac{\mathrm{i}}{2} \right)^{L/2}
\sum_{n=0}^L \sqrt{\begin{pmatrix} L \\ n \end{pmatrix}} \ket{\Psi_n},
\end{split}
\end{equation}
which would exhibit perfect revivals when evolved by $H_2$. On the other hand, the state $\ket{\mathbb{Z}_3}$ is expected to be a thermal state.

\begin{figure}[tbp]
    \centering
    \includegraphics[width=0.9\linewidth]{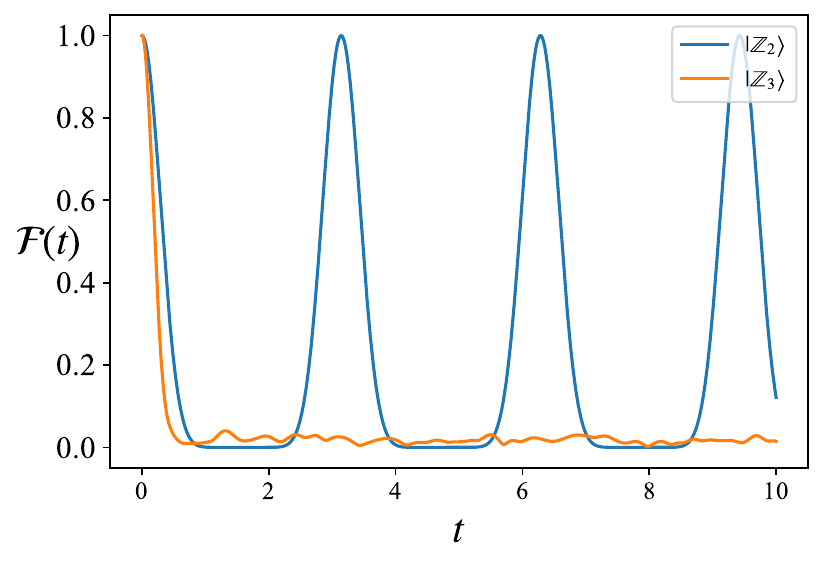}
    \caption{Fidelity dynamics for the Hamiltonian $H_2(g = 1, h_y =1)$ [Eq. \eqref{eq:XYZ_H}] with $(J_x, J_y, J_z) = (0.1, 0.5, 1.0)$ and $L=12$. Each $c_j$ is randomly chosen from $[-1, 1]$. 
    Periodic revivals can be seen when the initial state is $\ket{\mathbb{Z}_2}$ (blue), whereas for $\ket{\mathbb{Z}_3}$ (orange) the fidelity decays rapidly to $0$.}
    \label{fig:fidelity}
\end{figure}

In order to confirm this expectation, we calculate the time evolution of the fidelity $\mathcal{F}(t)\coloneqq\abs{\braket{\phi(t=0)}{\phi(t)}} = \abs{\mel{\phi}{e^{-\iu H_2t}}{\phi}}$. Figure \ref{fig:fidelity} shows the numerical results for the fidelity of $\ket{\mathbb{Z}_2}$ and that of $\ket{\mathbb{Z}_3}$. 
It is clear that the N{\' e}el state shows perfectly periodic revivals, which implies that it never reaches thermal equilibrium. We note in passing that a perfect revival of the initial state after a time of at most ${\cal O}(\mathrm{poly}(L))$, in general, implies the existence of QMBS \cite{alhambra2020revival}.

Since $\ket{\mathbb{Z}_2} = \ket{\uparrow\downarrow\uparrow\downarrow\cdots}$ is a zero-energy eigenstate of $H_2(g, h_y=0)$, we obtain the explicit form of the fidelity of this state:
\begin{equation}
    \mathcal{F}(t) = \abs{\mel{\mathbb{Z}_2}{e^{-\mathrm{i} t H_2(g , h_y)}}{\mathbb{Z}_2}} = \cos^L(h_yt).
\end{equation}
This readily implies that the fidelity of $\ket{\mathbb{Z}_2}$ shows perfect revivals with period $T=\frac{\pi}{h_y}$, which agrees with the numerical result shown in Figure \ref{fig:fidelity}. 

\begin{figure}[t]
    \centering
    \includegraphics[width=0.9\linewidth]{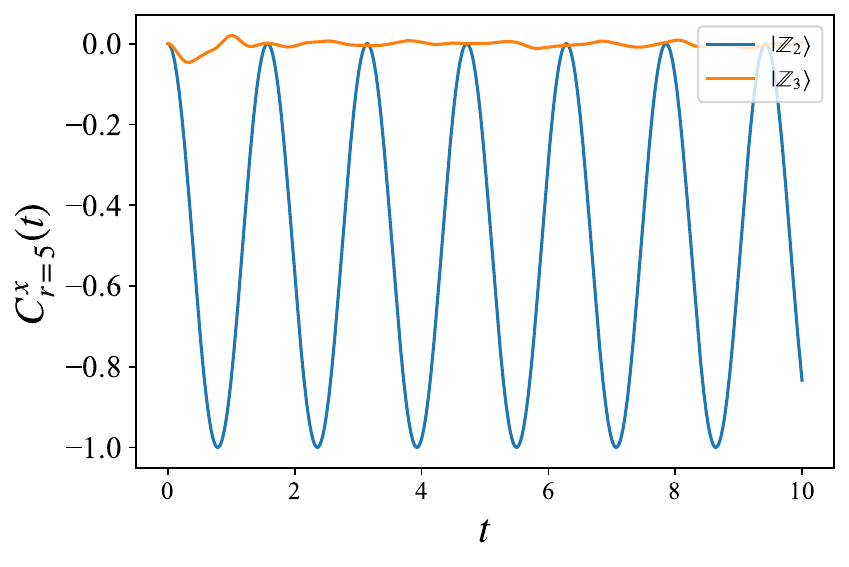}
    \caption{The time evolution of the correlation function $C_{r=5}^x(t)$ [Eq. \eqref{eq:correlation}] for $\ket{\mathbb{Z}_2}$ (blue) and $\ket{\mathbb{Z}_3}$ (orange). 
    The system size is $L=12$, and the parameters of Hamiltonian $H_2$ [Eq. \eqref{eq:XYZ_H}] are $g = 1$, $h_y = 1$ and $(J_x, J_y, J_z)=(0.1, 0.5, 1.0)$. Each $c_j$ is drawn from $[-1, 1]$ randomly.}
    \label{fig:correlation}
\end{figure}

To provide further evidence that $\ket{\mathbb{Z}_2}$ is a nonthermal state, we study the time evolution of the spin-spin correlation function: 
\begin{equation}\label{eq:correlation}
    C_r^x(t) = \frac{1}{L}\sum_{j=1}^L\mel{\phi(t)}{\sigma_j^x\sigma_{j+r}^x}{\phi(t)},
\end{equation}
where $\ket{\phi(t)}=e^{-\mathrm{i} t H_2}\ket{\phi}$ with $\ket{\phi}$ being the initial state. Specifically, we consider $\ket{\mathbb{Z}_2}$ and $\ket{\mathbb{Z}_3}$ as initial states. 
Figure \ref{fig:correlation} shows the numerical results for $C_r^x(t)$. 
We can see that $\ket{\mathbb{Z}_2}$ shows perfectly periodic revivals, while $\ket{\mathbb{Z}_3}$ shows no correlation. Since $\ket{\mathbb{Z}_2}$ is a superposition of exact QMBS, the correlation function $C_r^x(t)$ for this state can be calculated analytically. As a result, we obtain
\begin{equation}
    C_r^x(t) = (-1)^r\sin^2(2h_y t).
\end{equation}
From this, we immediately see that $C_r^x(t)$ shows perfect
revivals with period $T=\frac{\pi}{2h_y}$. This analytical result agrees with the numerical result shown in Fig.~\ref{fig:correlation}.

\section{Extension to higher-dimensional lattice models}\label{sec:higher-dimensional}
\begin{figure}[tbp]
    \centering
    \begin{tikzpicture}[scale=0.85]
        \draw (-0.2,-0.2) grid (4.2,4.2);
        \foreach \x in {0, ..., 4}
            \foreach \y in {0, ..., 4} {
                \fill (\x, \y) circle (2pt);
            }
        \draw (-0.2, 4.8) grid (4.2, 5.2);
        \draw (4.8, -0.2) grid (5.2, 4.2);
        \foreach \x in {0, ..., 5}{
            \draw[dotted] (\x, 4.2) -- (\x, 4.8);
            \draw[dotted] (\x, -0.2) --(\x, -0.8);
        }
        \foreach \y in {0, ..., 5}{
            \draw[dotted] (4.2, \y) -- (4.8, \y);
            \draw[dotted] (-0.2, \y) -- (-0.8, \y);
        }
        \draw[dotted, thick] (4.3, 4.3) -- (4.7, 4.7);
        \foreach \d in {0, ..., 4}{
            \fill (\d, 5) circle (2pt);
            \fill (5, \d) circle (2pt);
            \draw[dotted] (\d, 5.2) -- (\d, 5.8);
            \draw[dotted] (5.2, \d) -- (5.8, \d);
        }
        \draw[-stealth, thick] (-1.1, 0.5) to [edge label = $\bm{e}_y$] (-1.1, 1.5);
        \draw[-stealth, thick] (0.5, -1.1) to [edge label' = $\bm{e}_x$] (1.5, -1.1);
        \coordinate (O) at (0, 0) node [below left] at (O) {$(0, 0)$};
        \coordinate (Ly) at (0, 5) node [above left] at (Ly) {$(0, L_y) \equiv (0, 0)$};
        \coordinate (Lx) at (5, 0) node [below right] at (Lx) { $(L_x, 0) \equiv (0, 0)$ };

    \end{tikzpicture}
    \caption{An example of the $L_x \times L_y$ square lattice, where ${\bm e}_x=(1,0)$ and ${\bm e}_y=(0,1)$. Due to periodic boundary conditions, $(mL_x+x, nL_y+y)$ is identified with $(x, y)$ for any $m, n \in \mathbb{Z}$.}
    \label{fig:grid}
\end{figure}
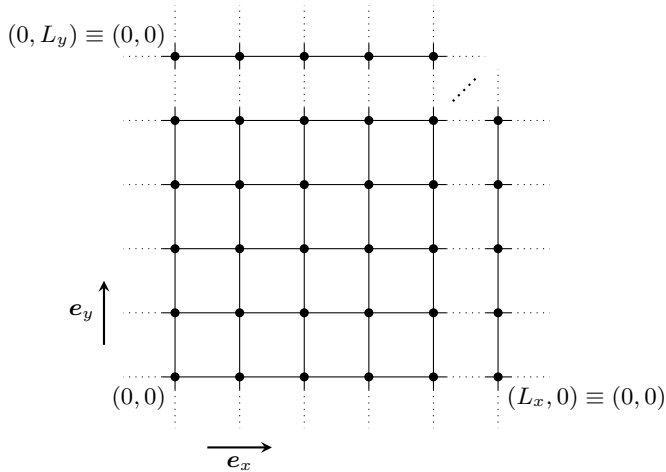

So far, our discussion has been limited to 1D models. One can, however, construct scarred models 
in higher dimensions in a similar manner. Let us consider a two-dimensional (2D) spin-$1/2$ model on the $L_x\times L_y$ square lattice $\Lambda$ with periodic boundary conditions (Fig.~\ref{fig:grid}), where both $L_x$ and $L_y$ are assumed to be even. We denote by $\sigma^\mu_{\bm j}$ ($\mu=x,y,z$) the Pauli matrices at site ${\bm j}=(j_x, j_y) \in \Lambda$. The Hamiltonian of the model is given by
\begin{equation}\label{eq:2d-ham}
    H^\mathrm{2d}(g , h_y) = Q_3^\mathrm{2d} + g H_\mathrm{pert}^\mathrm{2d} + h_y Y^\mathrm{2d},
\end{equation}
where
\begin{align}
    Q_3^\mathrm{2d} &= \sum_{\bm{j}\in\Lambda}\sum_{a=1}^3\sum_{b=a+1}^4  \lambda_{\bm{e}_b-\bm{e}_a} \hat{\bm{\sigma}}_{\bm{j}+\bm{e}_a}\cdot(\bar{\bm{\sigma}}_{\bm{j}} \times\hat{\bm{\sigma}}_{\bm{j}+\bm{e}_b}), \\
    H_\mathrm{pert}^\mathrm{2d} &= \sum_{\langle \bm{i}, \bm{j}\rangle}c_{\bm{i}, \bm{j}}(\sigma^x_{\bm{i}}\sigma^x_{\bm{j}}-\sigma_{\bm{i}}^y\sigma_{\bm{j}}^y+\sigma_{\bm{i}}^z\sigma_{\bm{j}}^z+1), \\
    Y^\mathrm{2d} &= \sum_{{\bm j}\in\Lambda}\sigma_{\bm j}^y,
\end{align} 
with $(\bm{e}_1, \bm{e}_2, \bm{e}_3, \bm{e}_4) = (-\bm{e}_x, -\bm{e}_y, \bm{e}_y, \bm{e}_x)$. See Fig. \ref{fig:grid} for the coordinate system and Eqs. \eqref{eq:hatspin} and \eqref{eq:barspin} for the definitions of $\hat{\bm{\sigma}}_{\bm{j}}$ and $\bar{\bm{\sigma}}_{\bm{j}}$, respectively. The interaction in the three-body term $Q_3^\mathrm{2d}$ takes the form of Eq. \eqref{eq:XYZQ3}. The prefactors are chosen to be $(\lambda_{2\bm{e}_x}, \lambda_{2\bm{e}_y}, \lambda_{\bm{e}_x+\bm{e}_y},\lambda_{\bm{e}_x-\bm{e}_y}) = (\lambda_1, \lambda_2, \lambda_3, \lambda_4)$, which can be arbitrary real numbers. 
The prefactors $c_{\bm{i}, \bm{j}}$ in the second term are arbitrary real numbers. 
Schematically, we can express $Q_3^\mathrm{2d}$ using the following diagrams,
\begin{widetext}
\begin{align}
    Q_3^\mathrm{2d} = \sum_{\bm{j} \in \Lambda}\left(
    \lambda_1\, 
    \begin{tikzpicture}[decoration={
    markings,
    mark=at position 0.65 with {\arrow{latex}}}, baseline= (j.base), scale=0.7]
        \coordinate[label = 225:$\bm{j}$] (j) at (1, 1);
        \draw[postaction={decorate}, thick, red] (0, 1) -- (1, 1); 
        \draw[postaction={decorate}, thick, red] (1, 1) -- (2, 1);
        \draw[dotted] (1,0) -- (1, 1) -- (1, 2);
        \fill (j) circle (2.5pt);
        \foreach \x in {0, 2}{
            \fill (\x, 1) circle (2pt);
            \fill (1, \x) circle (2pt);
        }
    \end{tikzpicture} + \lambda_2\,
    \begin{tikzpicture}[decoration={
    markings,
    mark=at position 0.65 with {\arrow{latex}}}, baseline= (j.base), scale=0.7]
        \coordinate[label = 225:$\bm{j}$] (j) at (1, 1);
        \draw[postaction={decorate}, thick, red] (1, 0) -- (1, 1); 
        \draw[postaction={decorate}, thick, red] (1, 1) -- (1, 2);
        \draw[dotted] (0,1) -- (1, 1) -- (2, 1);
        \fill (j) circle (2.5pt);
        \foreach \x in {0, 2}{
            \fill (\x, 1) circle (2pt);
            \fill (1, \x) circle (2pt);
        }
    \end{tikzpicture} + \lambda_3\,
    \begin{tikzpicture}[decoration={
    markings,
    mark=at position 0.65 with {\arrow{latex}}}, baseline= (j.base), scale=0.7]
        \coordinate[label = 225:$\bm{j}$] (j) at (1, 1);
        \draw[postaction={decorate}, thick, red] (0, 1) -- (1, 1); 
        \draw[postaction={decorate}, thick, red] (1, 1) -- (1, 2);
        \draw[dotted] (1,0) -- (1, 1) -- (2, 1);
        \fill (j) circle (2.5pt);
        \foreach \x in {0, 2}{
            \fill (\x, 1) circle (2pt);
            \fill (1, \x) circle (2pt);
        }
    \end{tikzpicture} + \lambda_3\,
    \begin{tikzpicture}[decoration={
    markings,
    mark=at position 0.65 with {\arrow{latex}}}, baseline= (j.base), scale=0.7]
        \coordinate[label = 225:$\bm{j}$] (j) at (1, 1);
        \draw[postaction={decorate}, thick, red] (1, 0) -- (1, 1); 
        \draw[postaction={decorate}, thick, red] (1, 1) -- (2, 1);
        \draw[dotted] (0,1) -- (1, 1) -- (1, 2);
        \fill (j) circle (2.5pt);
        \foreach \x in {0, 2}{
            \fill (\x, 1) circle (2pt);
            \fill (1, \x) circle (2pt);
        }
    \end{tikzpicture} + \lambda_4\,
    \begin{tikzpicture}[decoration={
    markings,
    mark=at position 0.65 with {\arrow{latex}}}, baseline= (j.base), scale=0.7]
        \coordinate[label = 225:$\bm{j}$] (j) at (1, 1);
        \draw[postaction={decorate}, thick, red] (1, 2) -- (1, 1); 
        \draw[postaction={decorate}, thick, red] (1, 1) -- (2, 1);
        \draw[dotted] (1,0) -- (1, 1) -- (0, 1);
        \fill (j) circle (2.5pt);
        \foreach \x in {0, 2}{
            \fill (\x, 1) circle (2pt);
            \fill (1, \x) circle (2pt);
        }
    \end{tikzpicture} + \lambda_4\,
    \begin{tikzpicture}[decoration={
    markings,
    mark=at position 0.65 with {\arrow{latex}}}, baseline= (j.base), scale=0.7]
        \coordinate[label = 225:$\bm{j}$] (j) at (1, 1);
        \draw[postaction={decorate}, thick, red] (0, 1) -- (1, 1); 
        \draw[postaction={decorate}, thick, red] (1, 1) -- (1, 0);
        \draw[dotted] (2,1) -- (1, 1) -- (1, 2);
        \fill (j) circle (2.5pt);
        \foreach \x in {0, 2}{
            \fill (\x, 1) circle (2pt);
            \fill (1, \x) circle (2pt);
        }
    \end{tikzpicture} 
    \right),
    \label{eq:Q32d}
\end{align}
with 
\begin{equation}\label{eq:fig_q3_pert}
    \begin{tikzpicture}[decoration={
    markings,
    mark=at position 0.65 with {\arrow{latex}}}, baseline= (j.base), scale=0.7]
        \coordinate[label = below:$\bm{i}$] (i) at (0, 0);
        \coordinate[label = below:$\bm{j}$] (j) at (1, 0);
        \coordinate[label = below:$\bm{k}$] (k) at (2, 0);
        \draw[postaction={decorate}, thick, red] (i) -- (j); 
        \draw[postaction={decorate}, thick, red] (j) -- (k);
        \foreach \x in {0, 1, 2}{
            \fill (\x, 0) circle (2pt);
        }
    \end{tikzpicture} 
    \coloneqq \hat{\bm{\sigma}}_{\bm{i}}\cdot(\bar{\bm{\sigma}}_{\bm{j}} \times\hat{\bm{\sigma}}_{\bm{k}}).
\end{equation}

We can decompose the 2D Hamiltonian $Q_3^\mathrm{2d}$ into a sum of 1D Hamiltonians defined on different paths on the 2D square lattice. This can be illustrated as
\begin{equation}
\begin{split}
    Q_3^\mathrm{2d} \,\, =  \,\, & \lambda_1 \, \begin{tikzpicture}[baseline = (O.base), scale = 0.7]
        \coordinate (O) at (0, 1.4);
        \draw (0, 0) grid (3.5, 3.5);
        \foreach \y in {0, 1, 2, 3} {
            \draw[very thick, red, dashed] (-0.5, \y) -- (0, \y);
            \draw[very thick, red] (0, \y) -- (3.5, \y);
            \draw[very thick, red, dashed] (3.5, \y) -- (4, \y);
        } \foreach \x in {0, 1, 2, 3} {
            \draw[dashed] (\x, -0.5) -- (\x, 0);
            \draw[dashed] (\x, 3.5) -- (\x, 4);
        }
    \end{tikzpicture} +  \lambda_2 \,  \begin{tikzpicture}[baseline = (O.base), scale = 0.7]
        \coordinate (O) at (0, 1.4);
        \draw (0, 0) grid (3.5, 3.5);
        \foreach \y in {0, 1, 2, 3} {
            \draw[dashed] (-0.5, \y) -- (0, \y);
            \draw[dashed] (3.5, \y) -- (4, \y);
        } \foreach \x in {0, 1, 2, 3} {
            \draw[very thick, red, dashed] (\x, -0.5) -- (\x, 0);
            \draw[very thick, red] (\x, 0) -- (\x, 3.5);
            \draw[very thick, red, dashed] (\x, 3.5) -- (\x, 4);
        }
    \end{tikzpicture} + 
    \lambda_3 \, 
    \begin{tikzpicture}[baseline = (O.base), scale = 0.7]
        \coordinate (O) at (0, 1.4);
        \draw (0, 0) grid (3.5, 3.5);
        \foreach \y in {0, 1, 2, 3} {
            \draw[dashed] (-0.5, \y) -- (0, \y);
            \draw[dashed] (3.5, \y) -- (4, \y);
        }
        \foreach \x in {0, 1, 2, 3} {
            \draw[dashed] (\x,-0.5) -- (\x,0);
            \draw[dashed] (\x,3.5) -- (\x,4);
        }
        \draw[very thick, dashed, red] (-0.5, 3) -- (0, 3);
        \draw[very thick, red] (0, 3) -- (0, 3.5);
        \draw[very thick, dashed, red] (0, 3.5) -- (0, 4);
        \draw[very thick, dashed, red] (-0.5, 1) -- (0, 1);
        \draw[very thick, red] (0, 1) -- (0, 2)--(1,2)--(1,3)--(2,3)--(2,3.5);
        \draw[very thick, dashed, red] (2, 3.5) -- (2, 4);
        \draw[very thick, dashed, red] (0,-0.5) -- (0, 0);
        \draw[very thick, red] (0, 0) -- (1, 0)--(1,1)--(2,1)--(2,2)--(3,2)--(3,3)--(3.5,3);
        \draw[very thick, dashed, red] (3.5,3) -- (4, 3);
        \draw[very thick, dashed, red] (2, -0.5) -- (2, 0);
        \draw[very thick, red] (2, 0)--(3,0)--(3,1)--(3.5,1);
        \draw[very thick, dashed, red] (3.5,1) -- (4, 1);
    \end{tikzpicture} \\
    + & \lambda_3 \, 
    \begin{tikzpicture}[baseline = (O.base), scale = 0.7]
        \coordinate (O) at (0, 1.4);
        \draw (0, 0) grid (3.5, 3.5);
        \foreach \y in {0, 1, 2, 3} {
            \draw[dashed] (-0.5, \y) -- (0, \y);
            \draw[dashed] (3.5, \y) -- (4, \y);
        }
        \foreach \x in {0, 1, 2, 3} {
            \draw[dashed] (\x,-0.5) -- (\x,0);
            \draw[dashed] (\x,3.5) -- (\x,4);
        }
        \draw[very thick, dashed, red] (-0.5, 2) -- (0, 2);
        \draw[very thick, red] (0, 2) -- (0, 3)--(1,3)--(1,3.5);
        \draw[very thick, dashed, red] (1, 3.5) -- (1, 4);
        \draw[very thick, dashed, red] (-0.5, 0) -- (0, 0) ;
        \draw[very thick, red] (0, 0) -- (0, 1)--(1,1)--(1,2)--(2,2)--(2,3)--(3,3)--(3,3.5);
        \draw[very thick, dashed, red] (3, 3.5) -- (3, 4);
        \draw[very thick, dashed, red] (1,-0.5) -- (1, 0);
        \draw[very thick, red] (1, 0) -- (2, 0)--(2,1)--(3,1)--(3,2)--(3.5,2);
        \draw[very thick, dashed, red] (3.5,2) -- (4,2);
        \draw[very thick, dashed, red] (3, -0.5) -- (3, 0);
        \draw[very thick, red] (3, 0)--(3.5,0);
        \draw[very thick, dashed, red] (3.5,0) -- (4,0);
    \end{tikzpicture} 
    + \lambda_4 \, 
    \begin{tikzpicture}[baseline = (O.base), scale = 0.7]
        \coordinate (O) at (0, 1.4);
        \draw (0, 0) grid (3.5, 3.5);
        \foreach \y in {0, 1, 2, 3} {
            \draw[dashed] (-0.5, \y) -- (0, \y);
            \draw[dashed] (3.5, \y) -- (4, \y);
        }
        \foreach \x in {0, 1, 2, 3} {
            \draw[dashed] (\x,-0.5) -- (\x,0);
            \draw[dashed] (\x,3.5) -- (\x,4);
        }
        \draw[very thick, dashed, red] (3.5,0) -- (4,0);
        \draw[very thick, red] (3.5,0) -- (3,0) -- (3,1)--(2,1)--(2,2)--(1,2)--(1,3)--(0,3)--(0,3.5);
        \draw[very thick, dashed, red] (0, 3.5) -- (0, 4);
        \draw[very thick, dashed, red] (3.5,2) -- (4,2);
        \draw[very thick, red] (3.5,2) -- (3,2) -- (3,3)--(2,3)--(2,3.5);
        \draw[very thick, dashed, red] (2, 3.5) -- (2, 4);
        \draw[very thick, red] (2,-0.5) -- (2,0) -- (1,0)--(1,1)--(0,1)--(0,2);
        \draw[very thick, dashed, red] (-0.5,2) -- (0,2);
        \draw[very thick, red] (0,-0.5) -- (0,0);
        \draw[very thick, dashed, red] (-0.5,0) -- (0,0);
    \end{tikzpicture} + \lambda_4 \, 
    \begin{tikzpicture}[baseline = (O.base), scale = 0.7]
        \coordinate (O) at (0, 1.4);
        \draw (0, 0) grid (3.5, 3.5);
        \foreach \y in {0, 1, 2, 3} {
            \draw[dashed] (-0.5, \y) -- (0, \y);
            \draw[dashed] (3.5, \y) -- (4, \y);
        }
        \foreach \x in {0, 1, 2, 3} {
            \draw[dashed] (\x,-0.5) -- (\x,0);
            \draw[dashed] (\x,3.5) -- (\x,4);
        }
        \draw[very thick, dashed, red] (3.5,1) -- (4,1);
        \draw[very thick, red] (3.5,1) -- (3,1) -- (3,2)--(2,2)--(2,3)--(1,3)--(1,3.5);
        \draw[very thick, dashed, red] (1,3.5) -- (1,4);
        \draw[very thick, dashed, red] (3.5,3) -- (4,3);
        \draw[very thick, red] (3.5,3) -- (3,3) -- (3,3.5);
        \draw[very thick, dashed, red] (3,3.5) -- (3,4);
        \draw[very thick, red] (3,-0.5) -- (3,0) -- (2,0)--(2,1)--(1,1)--(1,2)--(0,2)--(0,3);
        \draw[very thick, dashed, red] (-0.5,3) -- (0,3);
        \draw[very thick, red] (1,-0.5) -- (1,0)--(0,0)--(0,1);
        \draw[very thick, dashed, red] (-0.5,1) -- (0,1);
    \end{tikzpicture} \,\, ,
\end{split}\label{eq:1d_splitting}
\end{equation}
\end{widetext}

\noindent where each red line corresponds to the 1D Hamiltonian \eqref{eq:XYZQ3}. The real-valued parameters $\lambda_1$, $\lambda_2$, $\lambda_3$, and $\lambda_4$ can be chosen to be different. 
The above decomposition allows us to find the exact QMBS in the 2D model using the same way as in the 1D case.

Let us introduce the following state as a 2D analog of the tilted N\'{e}el states: 
\begin{align}
\ket{\psi^\mathrm{2d}_{1}(\alpha)} = \frac{1}{\sqrt{{\cal N}^\mathrm{2d}}} \bigotimes_{{\bm j} \in \Lambda} \ket{\chi}_{\bm j},
\end{align}
where
\begin{equation}
    \ket{\chi}_{\bm{j}} = \begin{cases}
        -\alpha\ket{\uparrow}_{\bm{j}}+\ket{\downarrow}_{\bm{j}} & j_x+j_y\equiv 0 \mod{2} \\
        \ket{\uparrow}_{\bm{j}}+\alpha\ket{\downarrow}_{\bm{j}} & j_x+j_y \equiv 1 \mod{2}
    \end{cases},
\end{equation}
and the normalization constant is given by ${\cal N}^\mathrm{2d}=(|\alpha|^2+1)^{N}$ with $N=L_x L_y$ denoting the total number of sites in $\Lambda$. The other tilted N\'{e}el state is obtained as
\begin{equation}
    \ket{\psi_2^\mathrm{2d}(\alpha)} = \mathcal{T}_x\ket{\psi_1^\mathrm{2d}(\alpha)} = \mathcal{T}_y\ket{\psi_1^\mathrm{2d}(\alpha)},
\end{equation}
where $\mathcal{T}_x$ and $\mathcal{T}_y$ are the translation operators along the $x$-axis and $y$-axis, respectively. Each of $\ket{\psi_{1,2}^\mathrm{2d}(\alpha)}$ can be regarded as an array of 1D tilted N\'{e}el states that reside on the red paths in each diagram of Eq. \eqref{eq:1d_splitting}. Thus, it is readily seen that these states are zero-energy states of $Q_3^\mathrm{2d}$. Similarly to the 1D case, they are annihilated by each local term in $H^\mathrm{2d}_\mathrm{pert}$. Therefore, the 2D tilted N\'{e}el states are zero-energy states of $H^\mathrm{2d}(g,0)=Q_3^\mathrm{2d}+g H^\mathrm{2d}_\mathrm{pert}$ for all $g$.

Similarly to the 1D case, we now consider the projections of $\ket{\psi_{1}^\mathrm{2d}(\alpha)}$ onto states with definite magnetization in the $y$ direction. They can be written in terms of the 2D version of the lowering operator as 
\begin{equation}
    \ket{{\Psi}^\mathrm{2d}_n} = \sqrt{\frac{(N-n)!}{N!\,n!}}
    \left( \mathcal{O}^-_\mathrm{2d} \right)^n \bigotimes_{{\bm j} \in \Lambda} |+\rangle_{\bm j} 
\end{equation}
where
\begin{equation}
    \mathcal{O}^-_\mathrm{2d} = \sum_{\bm{j}\in\Lambda} (-1)^{(j_x+j_y)}\Tilde{\sigma}^-_{\bm{j}}
\end{equation}
with $\Tilde{\sigma}^-_{\bm{j}}:=(\sigma^z_{\bm{j}}-\mathrm{i}\sigma^x_{\bm{j}})/2$, 
and $|+\rangle_{\bm j}$ is the eigenstate of $\sigma^y_{\bm j}$ with eigenvalue $+1$. 
Following the arguments in Sec. \ref{sec:XXX_based}, we find that $\ket{{\Psi}^\mathrm{2d}_n}$ is an eigenstate of the Hamiltonian (\ref{eq:2d-ham}) with eigenvalue $(N-2n)h_y$.

\begin{figure}
    \centering
    \includegraphics[width=0.97\linewidth]{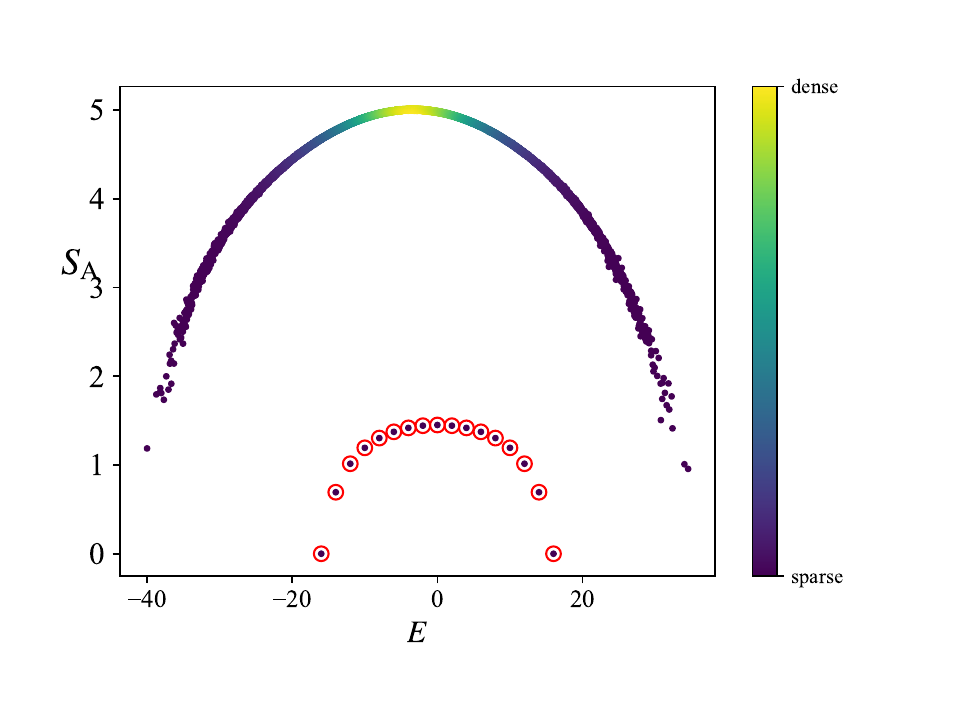}
    \caption{
    Half-system EE $S_A$ as a function of energy $E$ for all eigenstates of the Hamiltonian (\ref{eq:2d-ham}) with $(L_x, L_y) = (4, 4)$. The parameters of the Hamiltonian are $g=h_y =1$ and $(J_x, J_y, J_z) = (0.1, 0.5, 1.0)$. The random coefficients $c_{\bm{i}, \bm{j}}$ are chosen from an interval $[-1, 1]$.
    The density of data points is color coded. 
    The red solid circles indicate $\ket{\Psi^\mathrm{2d}_n}$. }
    \label{fig:2d_ee}
\end{figure}

To confirm that these states are low-entanglement states in the middle of the spectrum, we compute the EE for all energy eigenstates of the Hamiltonian (\ref{eq:2d-ham}). 
We divide the lattice $\Lambda$ into two halves $A$ and $B$ ($\Lambda=A \sqcup B$). Here, $A = \{\bm{j} \in \Lambda | j_y < L_y/2\}$ denotes the lower half, while $B$ denotes its complement. 
The analytical calculation of the half-system EE for QMBS, i.e., $S_A(\ket{{\Psi}^\mathrm{2d}_n})$, proceeds along the same lines as that of the ferromagnetic states discussed in~\cite{sanada2023quantum}. 
The result reads
\begin{equation}
    S_A(\ket{\Psi_n^\mathrm{2d}}) = -\sum_{k=0}^{\min({n, N-n})}\frac{\binom{N/2}{k}\binom{N/2}{n-k}}{ \binom{N}{n}}\ln \frac{\binom{N/2}{k}\binom{N/2}{n-k}}{\binom{N}{n}}.
    \label{eq:EE_2d_states}
\end{equation}
In particular, at $n = \frac{N}{2}$, it takes the maximum value, which is evaluated as 
\begin{equation}
    S_A\left(\ket{\Psi_{n=N/2}^\mathrm{2d}}\right) = \frac{1}{2}\ln \frac{e\pi N}{8} + O(N^{-1}).
\end{equation}
Thus, the entanglement entropy of $\ket{\Psi_n^\mathrm{2d}}$ obeys the sub-volume law, suggesting that these states are QMBS. Figure \ref{fig:2d_ee} shows the numerical results for the half-system EE for all eigenstates of $H^\mathrm{2d}(g , h_y)$. Clearly, the states $\ket{\Psi_n^\mathrm{2d}}$ are in the middle of the spectrum and stand out as entanglement outliers with extremely low entanglement. Their half-system EE perfectly match the analytical results \eqref{eq:EE_2d_states}. These results provide strong evidence that our 2D model is a scarred model with exact QMBS. 

Finally, we mention that our result can be generalized to models on any bipartite graph, for example, the honeycomb lattice. 
Here, a graph $G=(V, E)$ is bipartite if there exists a partition of $V$ into two disjoint sets
$A$ and $B$ such that every edge $e \in E$ connects a vertex in $A$ with one in $B$.
For a bipartite graph $G=(V, E)$, we introduce the following Hamiltonian 
\begin{equation}\label{eq:general_lattice}
    H^G(g, h_y) = Q^G + g H^G_\mathrm{pert} + h_y Y^G,
\end{equation}
with
\begin{align}
    Q^G &= \sum_{\langle ij \rangle, \langle jk \rangle \in E} \Hat{\bm{\sigma}}_{i}\cdot(\Bar{\bm{\sigma}}_{j} \times\Hat{\bm{\sigma}}_{k}), \\
    H^G_\mathrm{pert} &= \sum_{\langle ij \rangle \in E}c_{\langle ij \rangle}(\sigma_i^x\sigma_{j}^x-\sigma_i^y\sigma_{j}^y+\sigma_i^z\sigma_{j}^z+1), \\
    Y^G &= \sum_{v\in V} \sigma_v^y, 
\end{align}
where $c_{\langle ij \rangle}$ in the second term are arbitrary real numbers. 
Note that the three-spin term $\Hat{\bm{\sigma}}_{i}\cdot(\Bar{\bm{\sigma}}_{j} \times\Hat{\bm{\sigma}}_{k})$ vanishes identically for $i=k$. 
Similarly to the square-lattice case, one can define the lowering operator as
\begin{equation}
    \mathcal{O}_G^- = \sum_{v_a \in A} \Tilde{\sigma}_{v_a}^- -\sum_{v_b\in B}\Tilde{\sigma}_{v_b}^-.
\end{equation}
By repeatedly acting with this operator on the all-plus state, we can generate a tower of scar states
\begin{equation}
    \ket{\Psi^G_n} = (\mathcal{O}_G^-)^n \bigotimes_{v\in V}\ket{+}_v,
\end{equation}
which are exact eigenstates of $H^G(g, h_y)$. Through the same discussion as in the one-dimensional or 2D square-lattice cases, we can verify that these states violate ETH.

\section{Discussions}\label{sec:discussions}
In this article, we ahve constructed several models with multiple QMBS using IBS. We focused on the tilted-N\'{e}el states, which are parametrized IBS for the spin-$1/2$ Heisenberg and XYZ models, and showed that they serve as parent states of towers of scar states in a class of nonintegrable models. 
This is in contrast to our previous work, where the models constructed based on IBS exhibit only one or two isolated QMBS that do not admit any nontrivial dynamics such as periodic revivals. 
Thus, we have significantly improved our previous results, which indicates that the current method using IBS is as useful in constructing QMBS as other existing methods. 

Interestingly, the tower of QMBS we identified has an RSGA structure. The existence of an RSGA structure ensures that the QMBS generated by this method are equidistant in energy. We remark that, in a different context, examples of ETH-violating states with unequal level spacings has been constructed using partial solvability~\cite{Matsui_2024_1, PhysRevB.110.224306, wang2024generalizedspinhelixstates}. It would be interesting to extend our results in that context as well.

For future research, we aim to establish concrete relations between models with QMBS and integrable models. Through the construction of QMBS by our method, we reveal that some models with QMBS are closely related to integrable models via the IBS. Meanwhile, some integrable models share a common algebraic structure, such as the Temperley-Lieb algebra~\cite{temperley1971relations,baxter2016exactly}, chromatic algebra~\cite{fendley2010link, eck2024xxz}, or Birman-Murakami-Wenzl algebra~\cite{Kulish_2010, belavin2019algebraic}. 
Additionally, there are integrable models sharing the same algebraic structure but in different algebraic representations. Therefore, it would be worth constructing different ``representations'' of QMBS that emerge from IBS in different integrable models of the same algebraic origin. This intuition will provide valuable insights for developing a range of other models featuring QMBS.

\section*{Acknowledgements}
The numerical calculations were performed using QuSpin~\cite{weinberg2017quspin, weinberg2019quspin}. 
H.~K. is supported by JSPS KAKENHI Grants No.\ JP23K25783, No.\ JP23K25790, and MEXT KAKENHI Grant-in-Aid for Transformative Research Areas A “Extreme Universe” (KAKENHI Grant No.\ JP21H05191). 
K. S. acknowledges support from the Forefront Physics and Mathematics Program to Drive Transformation (FoPM). 
Y.M. would like to thank J. Minař and K. Schoutens for discussions, and the University of Amsterdam for hospitality. Y.M.'s work is supported by World Premier International Research Center Initiative (WPI), MEXT, Japan.

\section*{Data Availability}
The data supporting the findings of this study are available from the corresponding author upon reasonable request.
\appendix
\section{Proof that the tilted-N\'{e}el states are integrable boundary states}\label{appendix:NeelIBS}
In this appendix, we prove that the tilted-N\'{e}el states are IBS of the XYZ model. 
Let us fix $\eta, \tau\in\mathbb{C}$ with $\Im{\tau}>0$, which are the parameters of the elliptic functions in the following. 
Let $\mathbb{C}^2_j$ be the two-dimensional local Hilbert space at site $j$. 
The $R$-matrix of the XYZ model, acting on $\mathbb{C}^2_i \otimes \mathbb{C}^2_j$, is given as~\cite{Baxter_1971, Baxter_1973_1, Baxter_1973_2, Baxter_1973_3, baxter2016exactly, cao2014spin, slavnov20208vertex}
\begin{equation}
    R_{ij}(u) = \begin{pmatrix}
        a(u) & & & d(u) \\
        & b(u) & c(u) & \\
        & c(u) & b(u) & \\
        d(u) & & & a(u)
    \end{pmatrix},
\end{equation}
with 
\begin{align}
    a(u) &= \frac{\theta_4(\eta \vert 2\tau)\theta_1(u+\eta \vert 2\tau)}{\theta_1(\eta \vert 2\tau)\theta_4(u+\eta \vert 2\tau)}, \\
    b(u) &= \frac{\theta_4(\eta \vert 2\tau)\theta_1(u \vert 2\tau)}{\theta_1(\eta \vert 2\tau)\theta_4(u \vert 2\tau)}, \\
    c(u) &= 1, \\
    d(u) &= \frac{\theta_1(u+\eta \vert 2\tau)\theta_1(u \vert 2\tau)}{\theta_4(u+\eta \vert 2\tau)\theta_4(u \vert 2\tau)},
\end{align}
where $\theta_{n} (u | \tau)$ ($n\in \{1,2,3,4\}$) are the four Jacobi theta functions~\cite{slavnov20208vertex, gradshteyn2014table} with elliptic nome $q=e^{\mathrm{i} \tau}$. This $R$-matrix satisfies the regularity condition $R_{i, j}(u=0) = P_{i, j}$, where $P_{i, j}$ is a permutation operator such that $P_{i, j}(\ket{x}_i\otimes\ket{y}_j) = \ket{y}_i\otimes\ket{x}_j$ for any $\ket{x}, \ket{y} \in \mathbb{C}^2$.
The $R$-matrix satisfies the following Yang-Baxter equation (Fig. \ref{fig:yang-baxter}):
\begin{equation}\label{eq:yang-baxter}
\begin{split}
    R_{1,2}(u-v) & R_{1, 3}(u)R_{2, 3}(v) = \\
    & R_{2, 3}(v)R_{1, 3}(u)R_{1, 2}(u-v).
\end{split}
\end{equation}

\begin{figure}[tbp]
    \centering
    \begin{tikzpicture}[decoration={
    markings,
    mark=at position 0.65 with {\arrow{latex}}}, baseline= (r13.base), scale=0.7]
        \coordinate[label=200:$1$] (1s) at ({-0.6*sqrt(3)}, -0.6);
        \coordinate[label=below:$2$] (2s) at ({sqrt(3)}, -2.2);
        \coordinate[label=300:$3$] (3s) at ({1.6*sqrt(3)}, -1.6);
        \coordinate[label=above:$R_{1,3}$] (r13) at (0, 0);
        \coordinate[label=60:$R_{2, 3}$] (r23) at ({sqrt(3)}, -1);
        \coordinate[label=-60:$R_{1, 2}$] (r12) at ({sqrt(3)}, 1);
        \coordinate (1g) at ({1.6*sqrt(3)}, 1.6);
        \coordinate (2g) at ({sqrt (3)}, 2.2);
        \coordinate (3g) at ({-0.6*sqrt(3)}, 0.6);
        \draw[postaction={decorate}] (1s) --node[below] {$u$} (r13);
        \draw[postaction={decorate}] (r13) -- (r12);
        \draw[postaction={decorate}] (r12) --(1g);
        \draw[postaction={decorate}] (2s) -- node[left] {$v$}(r23);
        \draw[postaction={decorate}] (r23) -- (r12);
        \draw[postaction={decorate}] (r12) -- (2g);
        \draw[postaction={decorate}] (3s) -- (r23);
        \draw[postaction={decorate}] (r23) -- node[below] {$0$}(r13);
        \draw[postaction={decorate}] (r13) -- (3g);

        \draw ({1.6*sqrt(3)+1.5}, 0) node{$=$};

        \coordinate[label=240:$1$] (1s') at ({1.6*sqrt(3)+3}, -1.6);
        \coordinate[label=below:$2$] (2s') at ({2.2*sqrt(3)+3}, -2.2);
        \coordinate[label=300:$3$] (3s') at ({3.8*sqrt(3)+3}, -0.6);
        \coordinate[label=120:$R_{1, 2}$] (r12') at ({2.2*sqrt(3)+3}, -1);
        \coordinate[label=90:$R_{1, 3}$] (r13') at ({3.2*sqrt(3)+3}, 0);
        \coordinate[label=60:$R_{2, 3}$] (r23') at ({2.2*sqrt(3)+3}, 1);
        \coordinate (1g') at ({3.8*sqrt(3)+3}, 0.6);
        \coordinate (2g') at ({2.2*sqrt(3)+3}, 2.2);
        \coordinate (3g') at ({1.6*sqrt(3)+3}, 1.6);
        \draw[postaction={decorate}] (1s') --node[below] {$u$} (r12');
        \draw[postaction={decorate}] (r12') -- (r13');
        \draw[postaction={decorate}] (r13') -- (1g');
        \draw[postaction={decorate}] (2s') --node[right] {$v$} (r12');
        \draw[postaction={decorate}] (r12') -- (r23');
        \draw[postaction={decorate}] (r23') -- (2g');
        \draw[postaction={decorate}] (3s') -- node[below] {$0$} (r13');
        \draw[postaction={decorate}] (r13') -- (r23');
        \draw[postaction={decorate}] (r23') -- (3g');
    \end{tikzpicture}
    \caption{Graphical representation of the Yang-Baxter equation (Eq. \eqref{eq:yang-baxter}). We denote $R_{a, b}(u-v) = \protect\begin{tikzpicture}[decoration={
    markings,
    mark=at position 0.65 with {\protect\arrow{latex}}}, baseline = (O.base), scale=0.5]
        \protect\coordinate[label = left:$a$] (as) at (-1, 0);
        \protect\coordinate[label=below:$b$] (bs) at (0, -1);
        \protect\coordinate (O) at (0,-0.2);
        \protect\draw[postaction={decorate}] (as) -- node[above] {$u$}(0,0);
        \protect\draw[postaction={decorate}] (0,0) -- (1,0);
        \protect\draw[postaction={decorate}] (bs) -- node[right] {$v$}(0,0);
        \protect\draw[postaction={decorate}] (0,0) -- (0,1);
        \end{tikzpicture}$.  
        The diagrams are to be read from bottom to top, corresponding to reading \eqref{eq:yang-baxter} from right to left.
        }
    \label{fig:yang-baxter}
\end{figure}

From the $R$-matrix, we define the transfer matrix $T(u)$:
\begin{equation}
    T(u) = \Tr_a \left[ \overleftarrow{\prod_{j=1}^L} R_{a, j}\left(u-\frac{\eta}{2}\right)\right],
    \label{eq:defTu}
\end{equation}
where the multiplication is defined as $ \overleftarrow{\prod}_{j=1}^LA_j \coloneqq A_LA_{L-1}\cdots A_1$. 
Taking the logarithmic derivative of the transfer matrix,  we obtain the Hamiltonian and higher-order conserved charges, 
\begin{align}
    Q_n &= -\mathrm{i} \partial_u^{(n-1)} \log T(u)\vert_{u = 0}, \\
    H &\propto -\mathrm{i} \partial_u\log T(u)\vert_{u = 0} \propto Q_2.
\end{align}
Specifically, the anisotropy parameters of the XYZ Hamiltonian \eqref{eq:8vHam} are parametrized as
\begin{equation}
    J_x = \frac{\theta_4 (\eta \vert \tau )}{\theta_4 ( 0 \vert \tau )} , \, \, J_y = \frac{\theta_3 (\eta \vert \tau )}{\theta_3 ( 0 \vert \tau )} , \,\, J_z = \frac{\theta_2 (\eta \vert \tau )}{\theta_2 ( 0 \vert \tau )} .
\end{equation}

\begin{figure}[tbp]
\centering
\includegraphics[width = 0.95\linewidth]{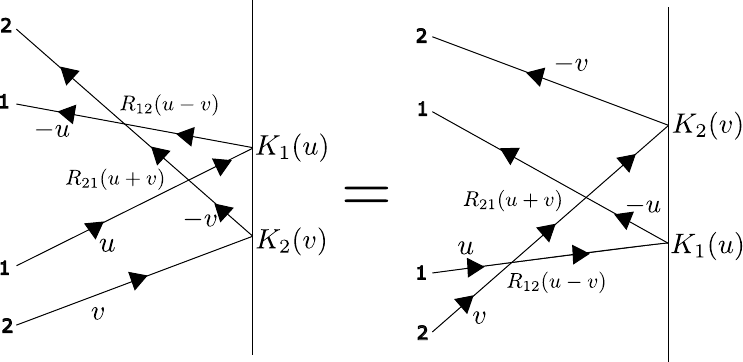}
\caption{Graphical representation of the boundary Yang-Baxter equation (Eq. \eqref{eq:boundary_yb_right}). The diagrams are to be read from bottom to top, corresponding to reading \eqref{eq:boundary_yb_right} from right to left. Note that the signs of the rapidities $u$ and $v$ change at the boundary because of the reflection.}
\label{fig:boundary_yb}
\end{figure}

Then we can translate the statement we want to prove into the language of the transfer matrix:
\begin{theorem}\label{thm:ibs_xyz}
    For $j = 1, 2$ and any $\alpha \in \mathbb{C}$, the following relation holds for tilted Néel states \eqref{eq:psi1} and \eqref{eq:psi2}, 
    \begin{equation}
        T(u)\ket{\psi_j(\alpha )} = \mathcal{I}T(u)\mathcal{I}\ket{\psi_j(\alpha )}, \,\, j \in \{1,2\} ,
    \end{equation}
    where $\mathcal{I}$ is the parity operator defined as
    \begin{equation}
        \mathcal{I} = \prod_{j=1}^{L/2} P_{j, L-j+1}.
    \end{equation}
\end{theorem}
The proof of this theorem is parallel to the proof that the tilted N\'{e}el states are IBS of the XXZ model~\cite{piroli2017integrable}. 
As outlined in \cite{piroli2017integrable}, the IBS can be viewed as integrable boundaries of the partition function of a vertex model, by rotating the products of transfer matrices by $\pi/2$. 
Therefore, we introduce the boundary Yang-Baxter equation~(BYBE) \cite{sklyanin1988boundary, deVega_1994, inami1994integrable, kimura2021bethe, piroli2017integrable} (Fig. \ref{fig:boundary_yb}), 
\begin{equation}
\begin{split}
    &R_{1, 2}(u-v)K_{R, 1}(u)R_{2, 1}(u+v)K_{R, 2}(v) \\
    &= K_{R, 2}(v)R_{2, 1}(u+v)K_{R, 1}(u)R_{1, 2}(u-v), \label{eq:boundary_yb_right}
    \end{split}
\end{equation}
\begin{equation}
\resizebox{.82\linewidth}{!}{$
\begin{aligned}
    &R_{1, 2}(-u+v)K_{L, 1}^t(u)R_{2, 1}(-u-v-2\eta)K_{L, 2}^t(v) \\
    &= K_{L, 2}^t(v)R_{2, 1}(-u-v-2\eta)K_{L, 1}^t(u)R_{1, 2}^t(-u+v),
\end{aligned}
    $}
\end{equation}
where the superscript $t$ denotes the transposition. The matrix $K_{R(L)}$ corresponds to the right (left) boundary condition~\cite{piroli2017integrable}.
We focus on the K matrices with non-vanishing off-diagonal elements, and we find solutions to the BYBE~\cite{inami1994integrable},
\begin{align}
    K_{R, a}(u; \xi, \mu, \lambda) &= \begin{pmatrix} K^{11}(u ) & K^{12}(u ) \\
        K^{21}(u ) & K^{22}(u )
    \end{pmatrix}_a, \\
    K_{L, a}(u) &= K_{R, a}^t(-u-\eta)   ,
\end{align}
where $\xi, \mu, \lambda \in\mathbb{C}$ is a free parameter. The matrix elements of the K matrix are
\begin{align}
    K^{11} (u) & = \frac{\theta_1(\xi + u \vert 2\tau)}{\theta_4(\xi + u \vert 2\tau)} , \,\, K^{22} (u) = \frac{\theta_1(\xi - u \vert 2\tau)}{\theta_4(\xi - u \vert 2\tau)} , \\
    K^{12}(u) & = \mu \frac{\theta_1( 2u \vert 2\tau)}{\theta_4( 2u \vert 2\tau)} \frac{\theta_4(\xi \vert 2\tau)^2(\lambda \epsilon_- +\epsilon_+ )}{\epsilon_0} ,\\
    K^{21}(u) & = \mu \frac{\theta_1( 2u \vert 2\tau)}{\theta_4( 2u \vert 2\tau)} \frac{\theta_4(\xi \vert 2\tau)^2(\lambda \epsilon_- -\epsilon_+ )}{\epsilon_0} ,
\end{align}
where the parameters are
\begin{align}
    \epsilon_0 & = \theta_4(\xi \vert 2\tau)^2\theta_4(u \vert 2\tau)^2 -\theta_1(\xi \vert 2\tau)^2\theta_1(u \vert 2\tau)^2 , \\
    \epsilon_\pm & = \theta_4(u \vert 2\tau)^2 \pm \theta_1( u \vert 2\tau)^2.
\end{align}

\begin{figure}
    \centering
    \includegraphics[width=\linewidth]{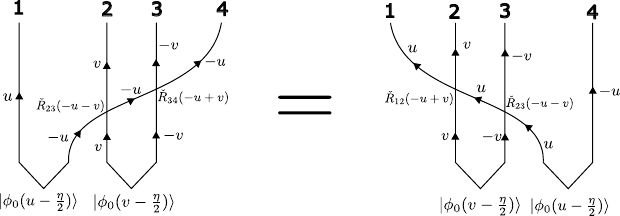}
    \caption{Graphical representation of the reflection equation (Eq. \eqref{eq:boundary_rotation}). The diagrams are to be read from bottom to top, corresponding to reading \eqref{eq:boundary_rotation} from right to left.}
    \label{fig:reflection_eq}
\end{figure}

By considering the $\pi/2$ rotation on the Euclidean spacetime, we obtain the reflection equation~\cite{piroli2017integrable}, as shown in Fig. \ref{fig:reflection_eq},
\begin{equation}\label{eq:boundary_rotation}
\resizebox{.82\linewidth}{!}{$
\begin{aligned}
    &\check{R}_{3, 4}(-u+v)\check{R}_{2,3}(-u-v)\ket{\phi_0(u-\frac{\eta}{2})}_{1,2}\ket{\phi_0(v-\frac{\eta}{2})}_{3, 4} \\ 
    &= \check{R}_{1, 2}(-u+v)\check{R}_{2, 3}(-u-v)\ket{\phi_0(v-\frac{\eta}{2})}_{1, 2}\ket{\phi_0(u-\frac{\eta}{2})}_{3, 4},
\end{aligned}
$}
\end{equation}   
where $\check{R}_{a, b}=R_{a, b}P_{a, b}$ and the boundary state is in the following form~\cite{piroli2017integrable},
\begin{equation}
\begin{split}
    \ket{\phi_0(u)}= & -K^{12}(u)\ket{\uparrow\uparrow}+K^{11}(u)\ket{\uparrow\downarrow} \\
    & -K^{22}(u)\ket{\downarrow\uparrow}+K^{21}(u)\ket{\downarrow\downarrow} .
\end{split}
\label{eq:phi0}
\end{equation}

\begin{figure}
    \centering
    \includegraphics[width=\linewidth]{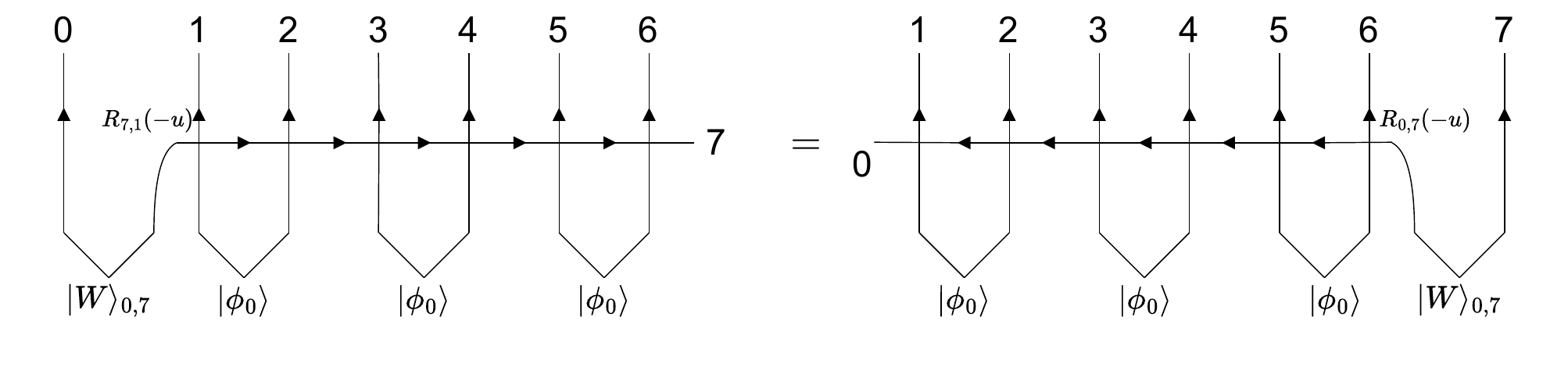}
    \caption{An illustration of Eq. \eqref{eq:reflection_eq_generalized} with $L=6$.}
    \label{fig:reflection_eq_generalized}
\end{figure}

We consider a boundary state with $(L+2)$ sites (labeled by $0$, $1$, $\cdots$, $L$, $(L+1)$) $|\phi_0(u-\frac{\eta}{2}) \rangle_{0,1} \otimes |\phi_0(v-\frac{\eta}{2}) \rangle_{2,3} \otimes \cdots \otimes |\phi_0(v-\frac{\eta}{2}) \rangle_{L,L+1} $ , as a generalization of the boundary states in Fig.~\ref{fig:reflection_eq}. Analogous to \eqref{eq:boundary_rotation}, and taking the limit $v \to 0$, we obtain
\begin{equation}
\begin{split}
    &\overleftarrow{\prod_{j=1}^L}R_{L+1, j}(-u) \left(\ket{W}_{0, L+1}\otimes\ket{\phi_0}^{\otimes \frac{L}{2}}\right)  \\
    &= \overleftarrow{\prod_{j=1}^L}R_{0, L+1-j}(-u) \left(\ket{\phi_0}^{\otimes \frac{L}{2}}\otimes\ket{W}_{0, L+1}\right),
\end{split}
\label{eq:reflection_eq_generalized}
\end{equation}
where $\ket{\phi_0} = \ket{\phi_0(-\frac{\eta}{2})}$ and $\ket{W}_{0, L+1}= \ket{\phi_0(u-\frac{\eta}{2})}_{0,L+1}$. Equation~\eqref{eq:reflection_eq_generalized} is illustrated in Fig.~\ref{fig:reflection_eq_generalized}. The sites $0$ and $L+1$ are auxiliary sites corresponding to the left and right boundary, respectively.

Finally, by comparing the components of \eqref{eq:reflection_eq_generalized} \cite{piroli2017integrable}, we obtain
\begin{equation}
\begin{split}
    &\Tr_{L+1} \left[\overleftarrow{\prod_{j=1}^L}R_{L+1, j}(-u)\right]\ket{\phi_0}^{\otimes \frac{L}{2}}  \\
    & = \Tr_{0}\left[\overleftarrow{\prod_{j=1}^L}R_{0, L+1-j}(-u)\right]\ket{\phi_0}^{\otimes \frac{L}{2}}.
\end{split}
\end{equation}
From the definition of the transfer matrix \eqref{eq:defTu}, the equation above implies 
\begin{equation}
    T(u)\ket{\phi_0}^{\otimes \frac{L}{2}} = \mathcal{I}T(u)\mathcal{I}\ket{\phi_0}^{\otimes \frac{L}{2}},
\end{equation}
satisfying the IBS condition.

Taking the limit $\lambda \to \infty$ with $\lim_{\lambda \to \infty} \mu \lambda = \zeta \in O(1)$ in \eqref{eq:phi0}, and defining $\alpha = \sqrt{\frac{K^{11}(-\eta/2)}{K^{22}(-\eta/2)}}$, the boundary state reads
\begin{equation}
\begin{split}
    \ket{\phi_0}^{\otimes \frac{L}{2}} & \propto \Big( - \alpha^2 | \uparrow \downarrow \rangle  + | \downarrow \uparrow \rangle + \\
    & \zeta w (-|\uparrow \uparrow \rangle + | \downarrow \downarrow \rangle ) \Big)^{\otimes \frac{L}{2}} ,
\end{split}
\end{equation}
where the constant is
\begin{equation}
    w = \frac{\theta_1(-\eta \vert 2 \tau ) \theta_4(\xi \vert 2 \tau )^2}{\theta_4(-\eta \vert 2 \tau )} \left. \frac{\epsilon_-}{\epsilon_0} \right|_{u = - \frac{\eta}{2}} .
\end{equation} 

If we further require that $\zeta = \frac{\alpha}{w}$, we obtain
\begin{equation}
\begin{split}
    \ket{\phi_0}^{\otimes \frac{L}{2}} \propto & \Big[ - \alpha^2 | \uparrow \downarrow \rangle  + | \downarrow \uparrow \rangle \\
    & +\alpha (-|\uparrow \uparrow \rangle + | \downarrow \downarrow \rangle ) \Big]^{\otimes \frac{L}{2}} \\
    & \propto | \psi_1 (\alpha) \rangle ,
\end{split}
\end{equation}
i.e., the tilted Néel state. Similar results can be obtained for $| \psi_2 (\alpha) \rangle = \prod_{n=1}^{L-1} P_{n,n+1} | \psi_1 (\alpha) \rangle$. Therefore, the tilted Néel states are also IBS of the XYZ model, proving Theorem \ref{thm:ibs_xyz}.

\bibliography{ref}

\end{document}